\documentclass
[superscriptaddress,secnumarabic,amssymb,amsmath,nobibnotes,aps,prd,showkeys,showpacs,nofootinbib,onecolumn]{revtex4}%
\usepackage{graphicx}
\usepackage{epsf}
\usepackage{bm}
\usepackage{amsmath}
\usepackage{amsfonts}
\usepackage{amssymb}%
\providecommand{\U}[1]{\protect\rule{.1in}{.1in}}
\newcommand{\be}{\begin{equation}}
\newcommand{\ee}{\end{equation}}

\newcommand{\mincir}{\raise
-3.truept\hbox{\rlap{\hbox{$\sim$}}\raise4.truept\hbox{$<$}\ }}
\newcommand{\magcir}{\raise
-3.truept\hbox{\rlap{\hbox{$\sim$}}\raise4.truept\hbox{$>$}\ }}

\begin{document}
\title{Symmetries and Singularities of the Szekeres System}
\author{Andronikos Paliathanasis}
\email{anpaliat@phys.uoa.gr}
\affiliation{Instituto de Ciencias F\'{\i}sicas y Matem\'{a}ticas, Universidad Austral de
Chile, Valdivia, Chile}
\affiliation{Institute of Systems Science, Durban University of Technology, POB 1334 Durban
4000, South Africa.}
\author{P.G.L. Leach}
\email{leach.peter@ucy.ac.cy}
\affiliation{Department of Mathematics and Institute of Systems Science, Research and
Postgraduate Support, Durban University of Technology, POB 1334 Durban 4000,
Republic of South Africa}
\affiliation{School of Mathematics, Statistics and Computer Science, University of
KwaZulu-Natal, Private Bag X54001, Durban 4000, Republic of South Africa}

\begin{abstract}
The Szekeres system is studied with two methods for the determination of
conservation laws. Specifically we apply the theory of group invariant
transformations and the method of singularity analysis. We show that the
Szekeres system admits a Lagrangian and the conservation laws that we find can
be derived by the application of Noether's theorem. The stability for the
special solutions of the Szekeres system is studied and it is related with the
with the Left or Right Painlev\'{e} Series which describes the expansions.

\end{abstract}
\keywords{Szekeres system; Silent universe; Lie symmetries; Singularity analysis}\maketitle
\date{\today}

\bigskip

Gravitational models with vanishing magnetic part of Weyl curvature tensor and
irrotational dust fluid component have been studied by Bruni, Matarrese and
Pantano in \cite{silent1}. Because there is no information dissemination with
gravitational or sound waves between the worldlines of neighboring fluid
elements these models are called silent universe \cite{silent2}. The field
equations form a system of six first-order ordinary differential equations.
The exact solutions at the critical points were found to be Tolman-Bondi,
Kantowski-Sachs or Szekeres geometries which can be seen as perturbations of
Friedmann-Lema\^{\i}tre-Robertson-Walker spacetimes \cite{silent1}. The
stability of these solutions was studied in \cite{silent3} where it was found
that solutions which describe pancakelike collapse are stable while those that
describe spindlelike collapse are unstable, while in \cite{musta} was shown
that the spacetimes that follows from the Szekeres system are "partially"
locally rotational spacetimes (PLRS). The integrability conditions for the
silent universes where studied in \cite{silent2} wherein a conjecture was
given such that "there are no spatially inhomogeneous irrotational dust silent
models, whose Weyl curvature tensor is of algebraic Petrov type I".

The field equations of the silent universe comprise a system of six
first-order ordinary differential equations\footnote{The system is an
algebraic-differential system in which the algebraic equation is related with
the curvature of the three-dimensional space.} in which the dependent
variables are: the energy density of the dust fluid,~$\rho$, the expansion
rate of the observer,~$\theta$, the shear components of the observer,
$\sigma_{1}$ and $\sigma_{2}$, and the two components of the electric part of
the Weyl curvature tensor,~i.e.~$E_{1}$ and $E_{2}$. A special case is when
the electric parts of the Weyl tensor are equal, i.e., $E_{1}=E_{2}$ as also
the shear $\sigma_{1}=\sigma_{2}$. In that case any irrotational dust models
with vanishing magnetic field is described by the so-called Szekeres system
\cite{szek0,barn1}. The dynamics of the Szekeres system can be found in
\cite{szek2}, while a covariant formulation based on the 1+1+2 decomposition
was derived recently in \cite{szek1}.

The equations which form the Szekeres system are%
\begin{align}
\dot{\rho}+\theta\rho &  =0,~\label{ss.01}\\
\dot{\theta}+\frac{\theta^{2}}{3}+6\sigma^{2}+\frac{1}{2}\rho &
=0,\label{ss.02}\\
\dot{\sigma}-\sigma^{2}+\frac{2}{3}\theta\sigma+\mathcal{E} &
=0,\label{ss.03}\\
\mathcal{\dot{E}}+3\mathcal{E}\sigma+\theta\mathcal{E}+\frac{1}{2}\rho\sigma
&  =0,\label{ss.04}%
\end{align}
plus the algebraic equation
\begin{equation}
\frac{\theta^{2}}{3}-3\sigma^{2}+\frac{^{\left(  3\right)  }R}{2}%
=\rho.\label{ss.00}%
\end{equation}
where a dot means contraction of the covariant derivative $\nabla_{\mu}$, with
respect to the timelike four-vector field $u^{\mu}$, such that $\dot
{A}=A_{;\mu}u^{\mu}$. $~$However except from the above system the $1+3$
analysis of the field equations provides the spacelike constraints
\cite{lesame}%
\begin{equation}
h_{\mu}^{\nu}\sigma_{\nu;\alpha}^{\alpha}=\frac{2}{3}h_{\mu}^{\nu}\theta
_{;\nu}~~,~h_{\mu}^{\nu}E_{\nu;\alpha}^{\alpha}=\frac{1}{3}h_{\mu}^{\nu}%
\rho_{;\nu}\label{ss.0a}%
\end{equation}
where $h_{\mu\nu}$ is the decomposable tensor defined by the expression
$h_{\mu\nu}=g_{\mu\nu}-\frac{1}{u_{\lambda}u^{\lambda}}u_{\mu}u_{\nu}$, and
$E_{\nu}^{\mu}=$ $Ee_{\nu}^{\mu},~\sigma_{\nu}^{\mu}=\sigma e_{\nu}^{\mu}$.
\ Obviously when the equations (\ref{ss.0a}) are satisfied identically then
equations (\ref{ss.01})-(\ref{ss.04}) describe a system of ordinary
differential equations, while in general they form a system of partial
differential equations. The spacelike constraint equations are essential for
the integrability of the silent models and in order the solution to reduce to
the Szekeres models \cite{silent2,lesame}. In the following we work with the
system (\ref{ss.01})-(\ref{ss.04}) and the constants of integration will be
functions independent on the time derivative which are constrained by the
system (\ref{ss.0a}).

The integrability of the Szekeres system (\ref{ss.01})-(\ref{ss.04}) has been
proved recently with the use of the Darboux polynomial and the Jacobi
multiplier methods \cite{szek}. Here we study the Szekeres system with two
different methods in the search for integrability and the existence of
analytical solutions. Specifically we use the symmetry method, group invariant
transformations and the singularity analysis. The purpose of this analysis is
to study the relationship between the two methods and see how the various
solutions, different universes, are related with the symmetries or the movable
singularities of the field equations\footnote{A discussion of these two
different methods \ for the study of integrability can be find in
\cite{aijgmmp}.}. Both methods have played an important role in gravitational
studies while recently they have been applied in modified theories of gravity
for the determinant of integrable field equations, for instance see
\cite{cap1,cap2,christ,vakili,palcqg,Demaret,Cotsakis,plaA} and references therein.

The jet-space which is defined by the Szekeres system is $J_{S}=\left\{
\tau,\rho,\theta,\sigma,\mathcal{E}\right\}  $ where $\tau$ is the independent
variable and it is a local coordinate of the spacetime. Consider the generator
$X$ of a one-parameter point transformation in space $J_{S}$. then
\begin{equation}
X=\xi\left(  \tau,Y^{B}\right)  \partial_{t}+\eta^{\rho}\left(  \tau
,Y^{B}\right)  \partial_{\rho}+\eta^{\theta}\left(  \tau,Y^{B}\right)
\partial_{\theta}+\eta^{\sigma}\left(  \tau,Y^{B}\right)  \partial_{\sigma
}+\eta^{\mathcal{E}}\left(  \tau,Y^{B}\right)  \partial_{\mathcal{E}%
},\label{ss.05}%
\end{equation}
where $Y^{B}=\left(  \rho,\theta,\sigma,\mathcal{E}\right)  $.

However, the Szekeres system can be written as a system of two second-order
ordinary differential equations. Without loss of generality we select the two
dependent variables to be the energy density, $\rho$, and the electric
component, $E$.

The system of second-order differential equations is
\begin{align}
&  \frac{d^{2}\rho}{d\tau^{2}}=\Phi\left(  \rho,\mathcal{E}\right)
\quad\mbox{\rm and}\\
&  \frac{d^{2}\mathcal{E}}{d\tau^{2}}=\Psi\left(  \rho,\mathcal{E}\right)
,\label{ss.06}%
\end{align}
where now the expansion rate and the shear are given from the relations%
\begin{align}
&  \theta=-\frac{\dot{\rho}}{\rho}\quad\mbox{\rm and}\\
&  \sigma=\frac{2\left(  \dot{\rho}\mathcal{E-}\rho\mathcal{\dot{E}}\right)
}{\rho\left(  \rho+6\mathcal{E}\right)  }.\label{ss.07}%
\end{align}
With the use of the latter expressions we observe that any point symmetry,
(\ref{ss.05}), of the Szekeres system is nothing else than a generalised
symmetry for the system, (\ref{ss.06}).

We now consider that the symmetry vector is the generator of a point
transformation in $\bar{J}_{S}=\left\{  \tau,\rho,\mathcal{E}\right\}  $. Then%
\begin{equation}
\bar{X}=\xi\left(  \tau,Y^{B}\right)  \partial_{t}+\eta^{\rho}\left(
\tau,Y^{B}\right)  \partial_{\rho}+\eta^{\mathcal{E}}\left(  \tau
,Y^{B}\right)  \partial_{\mathcal{E}}.\label{ss.08}%
\end{equation}
Note that $\bar{J}_{S}\subset J_{S}$, which means that by assuming a point
transformation at $\bar{J}_{S}$ we find only a partial number of symmetries
for the original system, but, as well see below, that it is sufficient in
order to prove the integrability of the system (\ref{ss.06}) as also to
determine an analytical solution. \ 

We find that the admitted Lie point symmetries of the second-order system
(\ref{ss.06}) are%
\[
X_{1}=\partial_{\tau},
\]%
\[
X_{2}=\frac{\rho\mathcal{E}}{\rho+6\mathcal{E}}\left(  6\partial_{\rho
}-\partial_{\mathcal{E}}\right)
\]
and
\[
X_{3}=\tau\partial_{\tau}-\frac{2\rho\left(  \rho+12\mathcal{E}\right)
\partial_{\rho}+12\mathcal{E}^{2}\partial_{\mathcal{E}}}{\rho+6\mathcal{E}}%
\]
which constitute the $A_{1}\otimes_{s}2A_{1}$ Lie algebra. $X_{1}$ is the
autonomous symmetry and $X_{3}$ is a rescaling symmetry which we see below
that the existence of both is important for the singularity analysis.

We apply the coordinate transformation\footnote{The new variables that we have
considered are related with the q-scalar variables which have been introduced
in \cite{szek2}. Specifically it holds $y^{3}=6\rho_{q}^{-1}$,~$x=\Delta
^{\left(  \rho\right)  }\left(  1+\Delta^{\left(  \rho\right)  }\right)
^{-1}$, where by definition it follows that $\theta_{q}=3\dot{y}y^{-1}$ and
$\dot{x}\left(  1-x\right)  ^{-1}=-3\sigma$.} to (\ref{ss.06}),%
\begin{equation}
\rho=\frac{6}{\left(  1-x\right)  y^{3}}~,~\mathcal{E}=\frac{x}{y^{3}\left(
x-1\right)  },\label{ss.09}%
\end{equation}
and the Szekeres system takes the simpler form,
\begin{equation}
\ddot{x}+2\frac{\dot{y}}{y}\dot{x}-\frac{3}{y^{3}}x=0,\label{ss.10}%
\end{equation}
and
\begin{equation}
\ddot{y}+\frac{1}{y^{2}}=0.\label{ss.11}%
\end{equation}

From (\ref{ss.11}) we find the conservation law%
\begin{equation}
I_{1}=\dot{y}^{2}-2y^{-1},\label{ss.12}%
\end{equation}
which gives
\[
\int\frac{dy}{\sqrt{I_{1}+2y^{-1}}}=\tau-\tau_{0}%
\]
while, when $I_{1}=0$, we have the closed-form solution
\[
y\left(  \tau\right)  =\left(  \frac{3\sqrt{2}}{2}\left(  \tau-\tau
_{0}\right)  \right)  ^{\frac{2}{3}}.
\]
Therefore, when we substitute $y\left(  \tau\right)  $ into (\ref{ss.10}), we
obtain a second-order linear equation which is maximally symmetric and
integrable. \ In the limit in which $I_{1}=0$, we find the closed-form
solution%
\begin{equation}
x\left(  t\right)  =x_{1}\left(  \tau-\tau_{0}\right)  ^{\frac{2}{3}}%
+x_{2}\left(  \tau-\tau_{0}\right)  ^{-1}.\label{ss.14}%
\end{equation}

Moreover we observe that the Szekeres system, (\ref{ss.10}), (\ref{ss.11}),
follows from the variation of the Action integral%
\begin{equation}
S=\int L\left(  x,\dot{x},y,\dot{y}\right)  d\tau\label{ss.15a}%
\end{equation}
where $L\left(  x,\dot{x},y,\dot{y}\right)  $ is point-like Lagrange function
\begin{equation}
L\left(  x,\dot{x},y,\dot{y}\right)  =y\dot{x}\dot{y}+x\dot{y}^{2}%
-xy^{-1}.\label{ss.15}%
\end{equation}
The Lagrangian is autonomous and is invariant under the action of the symmetry
vector $X_{1}$. Hence Noether's second theorem tells us that the corresponding
conservation law is the \textquotedblleft energy\textquotedblright,%
\begin{equation}
H\equiv\frac{p_{x}p_{y}}{y}-\frac{x}{y^{2}}\left(  p_{x}\right)  ^{2}+\frac
{x}{y}=h,\label{ss.16}%
\end{equation}
where $p_{x}$ and $p_{y}$ are the normal momenta, that is,%
\begin{equation}
y^{2}\dot{x}=yp_{y}-2xp_{x}~\mbox{\rm and}~y\dot{y}=p_{x}.\label{ss.17}%
\end{equation}

This is a Lagrange description of the field equation (\ref{ss.01}%
)-(\ref{ss.04}) for a specific lapse, derivative. However as it is well known
the field equations are invariant under the definition of another lapse i.e.
$d\tau\rightarrow N\left(  \tau^{\prime}\right)  d\tau^{\prime}$, where
$N\left(  \tau^{\prime}\right)  $ is an arbitrary function. However that it is
not true for the Action Integral (\ref{ss.15a}) and the corresponding
Euler-Lagrange equations will describe the Szekeres system. The description
that we have presented here will be used in order to consider the field
equations as equations of motion of a particle where methods of Classical
Mechanics can be studied. A discussion on the conformal equivalent Lagrangians
can be found in \cite{angrg}.

In terms of Noether's Theorem we can say that the conservation law, $I_{1}$,
is generated by a generalized symmetry and specifically a contact symmetry. It
holds that $\dot{I}_{1}=\left\{  I_{1},H\right\}  =0$, which means that the
dynamical system, (\ref{ss.10}), (\ref{ss.11}), is Liouville integrable. For
completeness we find that the system, (\ref{ss.10}), (\ref{ss.11},) admits
another linear-in-the-momentum conservation law which is nonautonomous and is
generated by the Lie symmetry vector $X_{3}$.

From (\ref{ss.16}) with the use of (\ref{ss.12}) we find that the solution of
the Hamilton-Jacobi Equation provides us with the Action%
\begin{equation}
S\left(  x,y\right)  =-\frac{\left(  I_{1}x+h\right)  }{I_{1}}\sqrt{\left(
I_{1}y^{2}+2y\right)  }-\frac{h}{\left(  I_{1}\right)  ^{\frac{3}{2}}}%
\ln\left(  \frac{I_{1}y+1}{\sqrt{I_{1}}}+\sqrt{\left(  I_{1}y^{2}+2y\right)
}\right)  \quad\mbox{\rm for}\quad I_{1}\neq0,\label{ss.18}%
\end{equation}
and%
\[
S\left(  x,y\right)  =\pm\sqrt{2y}\varepsilon x+hy^{\frac{2}{2}}%
\quad\mbox{\rm for}\quad I_{1}=0.
\]

From (\ref{ss.17}) the system of first-order ordinary differential equations
is%
\begin{equation}
\dot{x}=-\frac{\left(  I_{1}y+3\right)  x-hy}{y\sqrt{y\left(  I_{1}y+2\right)
}}\quad\\
\dot{y}=-\frac{\sqrt{y\left(  I_{1}y+2\right)  }}{y},\label{ss.19}%
\end{equation}
that is,%
\begin{equation}
\frac{dy}{dx}=\frac{\sqrt{y\left(  I_{1}y+2\right)  }}{\left(  I_{1}%
y+3\right)  x-hy}.\label{ss.20}%
\end{equation}
In the simplest scenario, $h=0$, it follows that $\int\sqrt{\frac{I_{1}y+2}%
{y}}dy=\ln x$. \ If we see the system (\ref{ss.01})-(\ref{ss.04}) as ordinary
differential equations then the latter solution corresponds to the Class II
\ solution of \cite{goode}. However in general the integration constants, and
the solution, are constrained by the spacelike constraints (\ref{ss.0a})

We now continue our analysis with the singularity analysis of the system,
(\ref{ss.10}, \ref{ss.11}). We apply the ARS algorithm which is described in
\cite{buntis}. We omit the details of the calculations. We find the
leading-order behaviours:%
\begin{equation}
x_{D}\left(  z\right)  =x_{0}z^{-1}~,~y_{D}\left(  z\right)  =y_{0}z^{\frac
{2}{3}},\label{ss.21}%
\end{equation}
and%
\begin{equation}
\bar{x}_{D}\left(  z\right)  =x_{0}z^{\frac{2}{3}}~,~\bar{y}_{D}\left(
z\right)  =y_{0}z^{\frac{2}{3}},\label{ss.22}%
\end{equation}
where $z=\tau-\tau_{0}$, $y_{0}$ satisfies the algebraic equation $y_{0}%
^{3}=\frac{9}{2}$ and $x_{0}$ is a constant of integration. All terms are
dominant. From (\ref{ss.14}) we can see that the leading-order powers solve
the field equations.

The resonances are found to be
\begin{equation}
s_{0}=-1~,~s_{1}=0~,~s_{2}=\frac{2}{3}~,~s_{3}=\frac{5}{3},\label{ss.23}%
\end{equation}
while for the leading-order behaviours in (\ref{ss.22}) the resonances are
\begin{equation}
\bar{s}_{0}=-1~,~\bar{s}_{1}=0~,~\bar{s}_{2}=\frac{2}{3}~,~\bar{s}_{3}%
=-\frac{5}{3}.\label{ss.24}%
\end{equation}
Note that the resonances are the same for every~ $y_{0}^{3}=\frac{9}{2}$. Last
but not least we performed the consistency test and we found that the field
equations pass the singularity test. That means that the solution can be
described in terms of a Laurent expansion.

The singularity analysis provides a different kind of integrability from that
of the symmetries. For instance, while the harmonic oscillator is a well-known
integrable system from the symmetry point of view, it fails to pass the
singularity test. However, as discussed in \cite{aijgmmp}, the singularity
analysis is coordinate dependent. It is important to realize that the
dependence of the results of the singularity analysis upon a coordinate
representation does not mean that the integrability in terms of singularity is
coordinate dependent. This means that for other coordinates maybe the solution
can be written in terms of a Laurent expansion.

From the resonances, (\ref{ss.23}), we have that the Laurent expansion is
given by a Right Painlev\'{e} Series, while a Left-Right Painlev\'{e} Series
describes the solution with resonances (\ref{ss.24}). However, the two
dominant terms and the two different sets of resonances do not mean that there
are different expansions which describe the solution. Actually it is only one
solution. That can be seen easily from the position of the constants of
integration for both the expansions which are the same. Moreover it is
important to mention that the dominant term $x_{D}$ is the special solution in
which $h=0$.

However, the special solutions, (\ref{ss.21}) and (\ref{ss.22}), are unstable,
that is, under a perturbation around the solutions or if the initial
conditions change such the values of $I_{1}$ and $h$ change, they fail to
describe the system. Of course that can be seen from the singularity analysis
because in the Right Painlev\'{e} Series the new terms which are given from
the expansion are becoming more dominant as we go further from the movable
singularity. However, in the special case in which the initial conditions
change such that $I_{1}$ continues to be zero, the special solution $\bar
{x}_{D}=x_{0}\tau^{2/3}$ continues to describe the system, which means that it
is stable. That property for this special solution is related with the
existence of the Left part for the Left-Right Painlev\'{e} Series.

Last but not least in order to relate the singularity analysis with the
symmetry vectors of the original system we want to say that the special
solution (\ref{ss.21}) is an invariant solution in the sense that it follows
from the zeroth-order invariants of the Lie symmetry vector $X_{3}$. These two
methods both provide the conservation law for the Szekeres system and show
that there is a direct equivalence of these two methods with the analysis
using the method of the Darboux polynomial and that of the Jacobi multiplier.
However, what it is more interesting is the existence of a point-like
Lagrangian which describes the evolution of the gravitational field equations,
while the conservation laws are Noetherian.

The two conservation laws with the use of the original coordinates~$\left(
\rho,\theta,\sigma,\mathcal{E}\right)  $ can be presented as the following
algebraic system%
\begin{equation}
I_{1}=\frac{2^{\frac{2}{3}}\left[  \left(  3\sigma+\theta\right)
^{2}-3\left(  \rho+6\mathcal{E}\right)  \right]  }{3^{\frac{4}{3}}\left(
\rho+6\mathcal{E}\right)  ^{\frac{2}{3}}}%
\end{equation}
and%
\begin{equation}
h=\left(  \frac{4}{3}\right)  ^{\frac{1}{3}}\frac{18\mathcal{E}^{2}%
+3\rho\sigma\left(  3\sigma+\theta\right)  +\mathcal{E}\left(  3\rho
-2\theta+6\sigma\left(  6\sigma+\theta\right)  \right)  }{\rho\left(
\rho+6\mathcal{E}\right)  ^{\frac{2}{3}}}%
\end{equation}
which defines the surfaces where the system evolves.

Furthermore,the Ricci scalar $R$ and the Weyl curvature invariant $\Psi_{2}$
are related with the variables $\left\{  x,y\right\}  $ as $x=6\frac{\Psi_{2}%
}{R}~,~y^{-3}=\Psi_{2}-\frac{1}{6}R~$where the homogeneous limit provides
$\left(  x,y\right)  \rightarrow\left(  0,0\right)  ~$\cite{szek2}. We can see
that the (special) power-law solutions (\ref{ss.21}), (\ref{ss.22}) do not
provide a homogeneous universe. 

In a forthcoming work we intend to study the quantization of the Szekeres
system by using the point-like Lagrangian that we found in the spirit of the
minisuperspace approach of quantum cosmology and also to extend our analysis
in the case of the silent universe.

\begin{acknowledgments}
We acknowledgement the anonymous referees for their comments and suggestions
which improved the quality of this work. AP acknowledges financial support
\ of FONDECYT grant no. 3160121. AP thanks the Durban University of Technology
for the hospitality provided while this work was performed. PGLL acknowledges
the National Research Foundation of South Africa and the University of
KwaZulu-Natal for financial support. The views expressed in this letter should
not be attributed to either institution.
\end{acknowledgments}

\end{document}